\documentclass[11pt]{article}

\usepackage{amsfonts,amsmath,amssymb,enumerate,graphicx}
\usepackage[normalem]{ulem}
\usepackage[usenames]{color}

\newcommand{\bpf}[1][Proof]{{\noindent {\sc #1: }}}
\newcommand{\epf}{{{\hfill $\Box$ \smallskip}}}

\newtheorem{theorem}{Theorem}

\newtheorem{remark}{Remark}

\newcommand{\Bc}{\mathcal{B}}
\newcommand{\Pp}{\mathsf{P}}
\newcommand{\R}{\mathbb{R}}
\newcommand{\N}{\mathbb{N}}

\newcommand{\Nc}{\mathcal{N}}
\newcommand{\Gc}{\mathcal{G}}
\newcommand{\Ic}{\mathcal{I}}
\newcommand{\Ec}{\mathcal{E}}

\newcommand{\ONE}{{\bf 1}}

\newcommand{\eps}{\varepsilon}

\newcommand{\var}{\mathop{\mathsf{Var}}}
\newcommand{\E}{\mathsf{E}}
\newcommand{\Law}{\mathop{\mathsf{Law}}}
\newcommand{\as}{\mathrm{a.s.}}

\title{Universal Statistics of Incubation Periods and Other Detection Times via Diffusion Models}
\author{Yuri Bakhtin\thanks{Courant Institute of Mathematical Sciences, New York}}

\begin{document}
\maketitle
\begin{abstract}
We suggest an explanation of typical incubation times statistical features based on the universal behavior of exit times for diffusion models. We give a
mathematically rigorous proof of the characteristic right skewness of the incubation time distribution for very general one-dimensional diffusion models. Imposing natural simple conditions on the drift coefficient, we also study these diffusion models under the assumption of  noise smallness and show that the limiting exit time distributions in the limit of vanishing noise are Gaussian and Gumbel. Thus they match the existing data as well as the other existing models do.   The character of our models, however, allows us to argue that the features of the exit time distributions that we describe are universal and manifest themselves in various other situations where the times involved can be described as detection or halting times, for example, response times studied in psychology.
\end{abstract}

\section{Introduction}
 Over the last hundred years, researchers accumulated a lot of data on incubation periods for various diseases in various populations.
 These data and the existing literature on the subject are thoroughly discussed in~\cite{Strogatz}, a recent paper that motivated the present study, so we are only
 giving a brief overview of the most imporant features of the data, referring to~\cite{Strogatz} and references therein for further details.
 
\begin{figure}
\includegraphics[width=12cm]{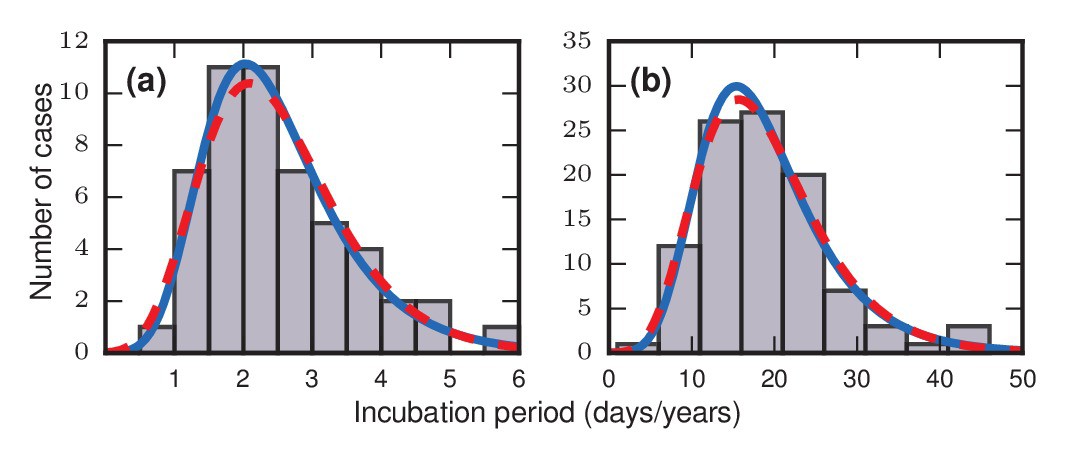}
\caption{\small Data redrawn from historic examples (reproduced from~\cite{Strogatz}, with the authors' permission).
Dashed red curves are lognormal densities and solid blue curves are Gumbel densities
predicted in~\cite{Strogatz}.
 (a) Data from an outbreak of food-borne streptococcal sore throat, reported in 
 \cite{SARTWELL:doi:10.1093/oxfordjournals.aje.a119397}, 
  time is measured in days. (b) Data from a study of bladder tumors among workers following occupational exposure to a carcinogen in a dye plant, \cite{Goldblatt65}. Time is measured in years. }
\label{fig:data-1949-and-1950}
\end{figure}

 All of these data show that
within the same population group, a simultaneous exposure to the same pathogen  does not result in simultaneous development of symptoms in
all individuals belonging to the group. Instead, those individuals who get sick show a broad distribution of incubation periods (i.e., times between the exposure and symptom onset). Moreover, the shapes
of observed distributions are strikingly similar to each other, being unimodal and right-skewed, with sharp decay on the left tail and extended decay on the right tail, see Figure~\ref{fig:data-1949-and-1950}.

Incubation  periods can be understood as the times needed for multiplication  of the harmful  agent
populations within the host organisms to reach a symptom onset threshold. The first
explanations of the right-skewness  were based on deterministic growth of the harmful 
agent population (such as exponential growth) with random parameters varying among individuals of the population and led to lognormal distribution of incubation times. However, there are
cases discussed in \cite{Strogatz} where the randomness in these parameteres is lacking but right-skewed distributions resembling lognormal are still observed. 

The new approach of~\cite{Strogatz} and a companion paper~\cite{Strogatz:PhysRevE.96.012313} is to model random incubation periods as stopping times for 
certain probabilistic models of the disease spread within an individual infected organism.  
In this approach, an organism is modeled by a network of nodes connected to each other by edges  and the
spread of the infection or disease is modeled by random evolution of labeling of the network nodes. Each node
is labeled as a healthy resident or a harmful invader, and then at each time step the label configuration is randomly updated according
to certain Markovian mechanism: a healthy resident with a harmful invader neighbor can randomly turn into a harmful invader and vice versa representing either reproduction or death of the disease agents.     
In addition to this, the network itself (its nodes and edges) may evolve according to a prescribed set of rules. The incubation period is modeled as a partial or complete takeover of the network by harmful invaders. 

Let us briefly summarize the findings of~\cite{Strogatz} without going into the details of  
the construction of this Markov process. The results depend on the network  geometry.
Massive computer simulations were carried out for several geometries and various values of parameters
of the Markov process involved such as the fitness of the harmful invaders. It was found numerically that
for all these situations the distribution of the time of complete or partial takeover 
is close either to Gumbel
or Gaussian distribution, depending on the details of the setup. 

In certain cases where the geometry of the network is simple
enough, precise limit theorems with Gumbel or Gaussian scaling limits 
in the infinite network size limit were obtained in~\cite{Strogatz} with mathematical rigor.
The Gumbel distribution is right-skewed and the Gaussian distribution is symmetric. In some cases, 
the evolution of the system was approximated by a simpler Markov chain also allowing for explicit computations that lead to a mathematical proof of positive skewness of the stopping time under conditioning on its finiteness. 
The authors are able to conclude that if the invader fitness is high then the model is similar to the classical ``coupon collection'' problem with right skew and limiting Gumbel distribution and if the fitness is low 
then the evolution of their model is similar to that a conditioned random walk that also results in the positive skew of the hitting time distribution.

The limitation of these results is that they are based on a very concrete model with specific update rules.  It is not obvious if the computations leading to the rigorous results in~\cite{Strogatz}
or the mathematical and numerical results themselves are valid for a broader class of models, and what precise conditions guarantee this or that kind of behavior.

The goal of the present paper is to suggest a broad class of models based on 
exit (or first passage) times for one-dimensional diffusions, i.e., Markov processes that solve
stochastic differential
equations (SDEs) on the real line. 
Despite the breadth, this class allows for rigorous analysis and precise mathematical statements on the random variables representing incubation times. An important advantage of our setup is that each SDE model comes with a whole universality class, i.e., a collection of discrete and continuous models that can be approximated by the SDE model. In fact, limit
theorems for stochastic processes with diffusion limits form a classical field of probability theory, see, e.g.,
\cite[Chapter~7]{Ethier-Kurtz:MR838085}.

Our main mathematical results are: 
\begin{enumerate}[(i)]
\item a mathematically rigorous proof of right-skewness for the exit time distribution conditioned on exit direction for our model under almost no assumptions besides the fact that it is a $1$-dimensional SDE (we also give a proof of right-skewness of exit times
for a general discrete nearest neighbor random walk aka birth-death process); 
\item  a description of limiting exit distributions (Gumbel or Gaussian) in the limit of vanishing noise, under natural simple assumptions on the drift of the diffusion process, based on existing rigorous mathematical results.   
\end{enumerate}

For incubation periods, this means that they are always right-skewed and that
they are approximately  either Gumbel or Gaussian, depending on the condition we impose.  Our results 
are stable with respect
to model modifications and thus describe large universality classes of systems whose macroscopic behavior
is insensitive to microscopic details. 

\section{Our model and main results}
\subsection{Modeling incubation periods with 1-dimensional SDEs}
\label{sec:setting}
We stress that we study the development of the disease within one infected individual and not the spread
of infection between individuals. In our mean field approach, we make a simplifying assumption that the state of the system describing the level of sickness in the individual at each time $t$ is represented by a single real variable $X(t)$. This variable may represent the size of the population of harmful invaders but may also be more involved. The values that $X(t)$ may take and that are of interest to us are concentrated on an interval $[0,R]$. Here,
the left endpoint $0$ corresponds to no sickness at all and represents an infection-free individual. The right endpoint $R$ is the level of the  disease corresponding to the onset of symptoms: we assume that the latent sickness develops unnoticed until it reaches the level~$R$.

We also assume that there is a point $x_0\in (0,R)$ such that the immune system of the infected individual does not detect the infection until the level of sickness reaches $x_0$. It is natural to assume that in many situations, in the absence of immune response, the time from the initial exposure to achieving the level~$x_0$ is  approximately constant (perhaps very close to $0$) and thus can be ignored in the study of the shape of the incubation period distribution. We further assume that after the immune system detects the infection,  $X$ is a time-homogeneous Markov process with continuous paths. Under broad conditions, such a process is a solution 
of an SDE:
\begin{equation}
\label{eq:basic_sde}
dX(t)= b(X(t))dt + \sigma(X(t))dW(t).
\end{equation}
The function $b(x)$  usually called the drift and assumed to be smooth in~$x$ represents the combined influence of the infection expansion and the immune response. These influences can be interpreted as the birth rate~$B(x)$ and death rate $D(x)$ of harmful invaders:
$b(x)=B(x)-D(x)$, $x\in [0,R]$.
The randomness in the system is modeled by white noise $dW$ in~\eqref{eq:basic_sde}, where~$W$ is a standard Wiener process or Brownian Motion. We denote probabilities of events  by $\Pp(\cdot)$. The smooth diffusion coefficient $\sigma(x)>0$ represents the amplitude of the noise at~$x\in [0,R]$.  

In our mean-field approach we assume that the SDE coefficients~$b$ and~$\sigma$ depend only on $x\in[0,R]$, the single state variable in the system, although more general setups are possible.  
We are going to model the incubation period by the exit time from $(0,R)$. Namely, we define the random variable $\tau_\eps$ as the first exit time for the process $X$ from~$(0,R)$:
\[
\tau=\inf\big\{t:X(t)=0\ \textrm{or}\ X(t)=R\big\}
\]
There are three possible outcomes of the evolution up to the exit time: 
\begin{enumerate}
\item $X(t)$ reaches $R$ before $0$,  i.e., $X(\tau)=R$. This means that the immune system was not successful in blocking the infection propagation, and at time $\tau$ the disease is strong enough for the symptom onset, so $\tau$ may be interpreted as the incubation time. The samples in all incubation time studies are based only on the individuals with this outcome.

\item $X(t)$ reaches $0$ before $R$,  i.e., $X(\tau)=0$. This means that the immune system has been succesful in complete elimination of the infection by time $\tau$ while no visible symptoms have ever developed. So $\tau$ can be interpreted as the latent disease healing time but the individuals that never develop any symptoms are not in the focus of this paper and the associated statistical data on infection elimination times is not available.

\item $X(t)$ never reaches endpoints $0$ or $R$ staying within $(0,R)$ for all times. In this case $\tau=+\infty$, and the latent infection persists indefinitely fluctuating above the zero level and never being detected. 
On the one hand, this situation has zero probability under our assumptions on the coefficients~$b$ and $\sigma$.
On the other hand, the individuals with such behavior are also excluded from incubation period statistical studies.   
\end{enumerate}

\subsection{Exit times conditioned on exit through a threshold are always right-skewed}

Our first result concerns the right-skewness of the exit time distribution conditioned on first exit through the right endpoint~$R$.
To state the theorem, we need some notation. Let us denote by $\Gamma$ the symptom onset event, i.e., $\Gamma= \{X(\tau)=R\}$. Under our assumptions, $\Pp(\Gamma)>0$ and conditioning on $\Gamma$ is well-defined. Under this conditioning, the exit time $\tau$ may be viewed as the first passage time for level~$R$.

The right-skewness of a distribution is formally defined via positivity of the skewness coeffeicient.  
Let us now recall the relevant definitions. For a random variable $Y$, its skewness $\gamma(Y)$ is defined by
\begin{equation}
\label{eq:def-of-skew}
    \gamma(Y) = \frac{\E(Y-\E Y)^3}{ \var(Y)^{3/2}}
              = \frac{\kappa_3(Y)}{\kappa_2^{3/2}(Y)}.
\end{equation}
Here $\E Y$ is the expectation of $Y$, $\var(Y)=\E(Y-\E Y)^2$ is the variance of~$Y$, and $\kappa_n(Y)$ stands for the $n$-th cumulant 
of~$Y$ defined by
\begin{equation*}
\kappa_n(Y) = \frac{1}{i^k} \left[\frac{d^n}{d\lambda^n} \ln \varphi_Y(\lambda) \right]_{\lambda=0}, 
\end{equation*}
where $\varphi_Y=\E e^{i\lambda Y}$ is the characteristic function of~$Y$, and $\ln$ denotes the main branch of the logarithm function. The cumulant $\kappa_k(Y)$ is well-defined if $\E |Y|^k<\infty$. Cumulants are Taylor coefficients for $\ln \varphi(\lambda)$ at~$0$:
\begin{equation*}
\ln \varphi_Y(\lambda)=\kappa_1 \frac{it}{1!}+\kappa_2\frac{(it)^2}{2!}+\ldots+\kappa_n\frac{(it)^n}{n!}+o(|t|^n), 
\end{equation*}
and can be expressed in terms of moments of $Y$. Denoting $\E Y^k =\alpha_k(Y)$, we have
\begin{equation}
\label{eq:cumulant-via-moments}
\kappa_n(Y)=\alpha_n(Y)+P(\alpha_1(Y),\dots,\alpha_{n-1}(Y)), 
\end{equation}
where $P$ is a polynomial with all monomials of degree at least $2$. The precise formula is given in, e.g., \cite[Section 2.12]{Shiryaev:MR1368405}. For $k=1,2,3,$
we have
\begin{align}
\kappa_1(Y)&=\E Y=\alpha_1(Y),\\
\kappa_2(Y)&= \var Y = \E (Y-\E Y)^2 =\E Y^2 - (\E Y)^2=\alpha_2(Y)-\alpha_1^2(Y) ,\\
\notag
\kappa_3(Y)&=\E (Y-\E Y)^3 = \E Y^3 -3 \E Y^2 \E Y + 2 (\E Y)^3 
\\&\hspace{4cm} = \alpha_3(Y) -3 \alpha_2(Y)\alpha_1(Y) + 2\alpha_1^3(Y).
\end{align}

If the moment generating function $M_Y(\lambda)=\E e^{\lambda Y}$ is defined for $\lambda$ in a neighborhood of~$0$, then 
\begin{equation*}
\kappa_n(Y) =  \left[\frac{d^n}{d\lambda^n} \ln M_Y(\lambda) \right]_{\lambda=0}, 
\end{equation*}
and
\begin{equation}
\label{eq:expansion_for_mgf}
\ln M_Y(\lambda)=\kappa_1 \frac{\lambda}{1!}+\kappa_2\frac{\lambda^2}{2!}+\ldots+\kappa_n\frac{\lambda^n}{n!}+\ldots 
\end{equation}

Our main result on skewness of exit times is:
\begin{theorem} 
\label{thm:right-skew}
Under the conditions described in Section~\ref{sec:setting}, conditioned on~$\Gamma$, \[\gamma(\tau)>0.\]
\end{theorem}

Due to \eqref{eq:def-of-skew}, this theorem is a direct consequence of the following:
\begin{theorem}
\label{thm:positive-cumulants}
Under the conditions described in Section~\ref{sec:setting}, conditioned on~$\Gamma$, 
\[
\kappa_n(\tau)>0,\quad n\in\N.
\]
\end{theorem}

These two theorems show that incubation periods are always right-skewed and, moreover,  all cumulants
of incubation periods are positive. We prove Theorem~\ref{thm:positive-cumulants} in Section~\ref{sec:proofs-of-positive-skewness}.
Although we do not estimate the magnitude of positive cumulants in this proof, such estimates are possible because the proof is based on a representation of $\kappa_n(\tau)$ as an integral of a positive quantity that can be estimated. 

Our proof of Theorem~\ref{thm:positive-cumulants} is direct but one could also derive it from the fact that under conditioning on~$\Gamma$, the distribution of $\tau$ is infinitely divisible and concentrated
on $[0,\infty)$. 
Infinite divisibility along with some other distributional properties of exit times of $1$-dimensional diffusions  conditioned on the direction of exit such as
 unimodality and log-concavity will be addressed in a separate publication.

Section~\ref{sec:proofs-of-positive-skewness} also contains proofs of versions of Theorems~\ref{thm:right-skew} and~\ref{thm:positive-cumulants}
for discrete random walks instead of continuous time SDEs.

\subsection{Exit time distributions in vanishing noise limit}
\label{sec:small_noise}
Next we study the situation where the deterministic effects dominate over the random ones in the disease development.
To formalize this, we consider a whole family of SDEs indexed by a small parameter $\eps>0$
and assume that $b$ does not depend on $\eps$ while $\sigma(x)=\sigma_\eps(x)=\eps\sigma_1(x)$ for some smooth function 
$\sigma_1(x)>0$. Then SDE~\eqref{eq:basic_sde} rewrites as
\begin{equation*}
dX_\eps(t)= b(X_\eps(t))dt + \eps\sigma_1(X_\eps(t))dW(t),
\end{equation*}
the solution and the associated exit time depend on $\eps$, and we denote them by $X_\eps$ and $\tau_\eps$. 

We will describe the limiting behavior of exit times as $\eps\to 0$. Generally speaking, dynamical systems under small noisy perturbations is a well-developed field, see, e.g., the classical monograph~\cite{FW2012}. 
Of course, the behavior of the SDE solutions depends crucially on the phase portrait of the vector field $b(x),x\in[0,R]$, i.e., on the structure of subsets of $[0,R]$ where
$b$ is positive, negative, and zero. We recall that a point $x$ is called critical for $b$ if $b(x)=0$.

We will consider the following three situations:
\begin{enumerate}[I.]
\item \label{it:right-drift} There are no critical points on $[0,R]$ and $b(x)>0$ for all $x\in[0,R]$.
\item \label{it:left-drift} There are no critical points on $[0,R]$ and $b(x)<0$ for all $x\in[0,R]$.
\item \label{it:barrier} There is exactly one critical point $p\in(0,R)$; $\lambda:=b'(p)>0$; $b(x)>0$ for all
$x\in(p,R]$; $b(x)<0$ for $x\in[0;p)$.
\end{enumerate}

In dimension~$1$, any vector field $b$ can be represented via gradient of a potential:
$b(x)=-\Phi(x)$. In cases \ref{it:right-drift} and \ref{it:left-drift}, $\Phi$ is monotone on $[0,R]$.
In case~\ref{it:barrier}, $\Phi$ has a maximum at $p$.

We will further subdivide Case~\ref{it:barrier} into two subcases: \ref{it:barrier}$_0$, where $x_0<p$,
and \ref{it:barrier}$_1$, where $x_0>p$. We ignore the exceptional case $p=x_0$ in this paper for brevity,
although the exit times have been studied for this case in detail starting with~\cite{Day:MR1110156}, 
more on this in Section~\ref{sec:discussion}.

\begin{figure}
\begin{center}
\includegraphics[height=3.5cm]{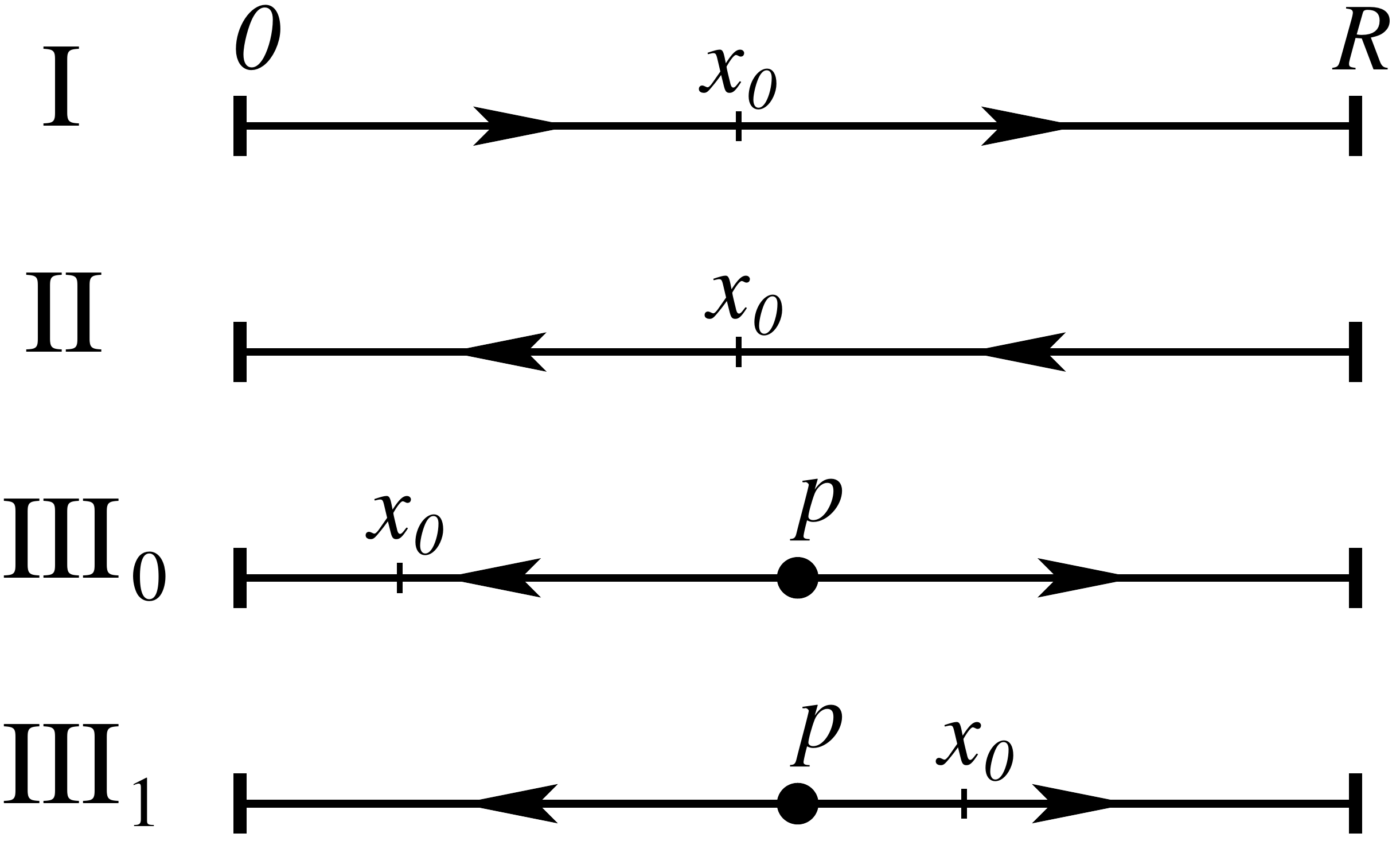}
\end{center}
\caption{\small The phase portraits considered in Section~\ref{sec:small_noise}}
\label{fig:portrait}
\end{figure}

The phase portraits for all these cases are given in Figure~\ref{fig:portrait}. 
The archetypal examples
of these cases are: 
\begin{equation}
\label{eq:canonical-examples}
b(x)=\begin{cases}
1,& \mathrm{case~\ref{it:right-drift}},\\
-1,& \mathrm{case~\ref{it:left-drift}},\\
\lambda (x-p),&  \mathrm{case~\ref{it:barrier}}.
\end{cases}
\end{equation}
In fact, for generic $b$ in each of the cases 
\ref{it:right-drift},\ref{it:left-drift},\ref{it:barrier}, there is a smooth coordinate change (conjugation)
$\tilde y=h(x)$ such that the motion along $b$  is transformed, in the new coordinates, into the motion along the 
associated canonical drift given in~\eqref{eq:canonical-examples}. In case~\ref{it:right-drift}, one simply can 
define $h(x)$ as the time it takes to travel from $0$ to $x$ along $b$;
case~\ref{it:left-drift} is similar; in case~\ref{it:barrier}, the conjugation is slightly more involved, 
see, e.g.,~\cite[Section 1]{Eizenberg:MR749377}.

The mathematical analysis of more sophisticated phase portraits is also possible but we consider these three simplest cases because they correspond to the following most natural situations: in 
case~\ref{it:right-drift}, the infection is stronger than the immune system over the entire interval~$[0,R]$; in case~\ref{it:left-drift}, the immune system is stronger than the infection propagation
over the entire interval~$[0,R]$; in case~\ref{it:barrier}, the immune system is stronger if the infection level is below the ``critical mass'' $p$, and if the infection level is above that critical mass, then the immune system is not strong enough to prevent the infection growth, at least in the regime described by the deterministic ODE $\dot x = b(x)$.

\smallskip

The symptom onset event describing an exit through the right endpoint depends on $\eps$ in this section, so we will denote it by
$\Gamma_\eps=\{X_\eps(\tau_\eps)=R\}$.

For small $\eps$, the event $\Gamma_\eps$  describes a typical outcome in Cases~\ref{it:right-drift} and~\ref{it:barrier}$_1$,
but it is a rare event in Cases~\ref{it:left-drift} and~\ref{it:barrier}$_0$. The precise mathematical meaning of this claim is given by
the following statement:
\begin{theorem}
\label{th:exit-0-1} In all the cases we are considering, $q=\lim_{\eps\to 0}\Pp(\Gamma_\eps)$ is well-defined. In cases~{\rm\ref{it:right-drift}} and~{\rm\ref{it:barrier}}$_1$, $q=1$; in cases~{\rm\ref{it:left-drift}} and~{\rm\ref{it:barrier}}$_0$,
$q=0$.
\end{theorem}

This theorem is a specific case of classical results on exit problems for small random perturbations of dynamical systems in the so called Levinson case (where the deterministic orbit started at $x_0$ hits the boundary), see~\cite[Section~2.1]{FW2012}. In all these cases the typical behavior consists in flowing along the vector field $b$ for a finite time.

\medskip

The notion of incubation period is valid only for individuals that develop symptoms, so for 
both types of limiting behavior of $\Pp(\Gamma_\eps)$ described by Theorem~\ref{th:exit-0-1}, we are interested in the statistics of $\tau_\eps$ conditioned on event~$\Gamma_\eps$.
We always have $\Pp(\Gamma_\eps)>0$, so for any random variable $Y$ its 
conditional distribution $\Law[\,Y\,|\,\Gamma_\eps]$   given that the first exit from $(0,R)$ 
happens through~$R$ is well-defined.

\begin{figure}
\begin{center}
\includegraphics[width=11cm]{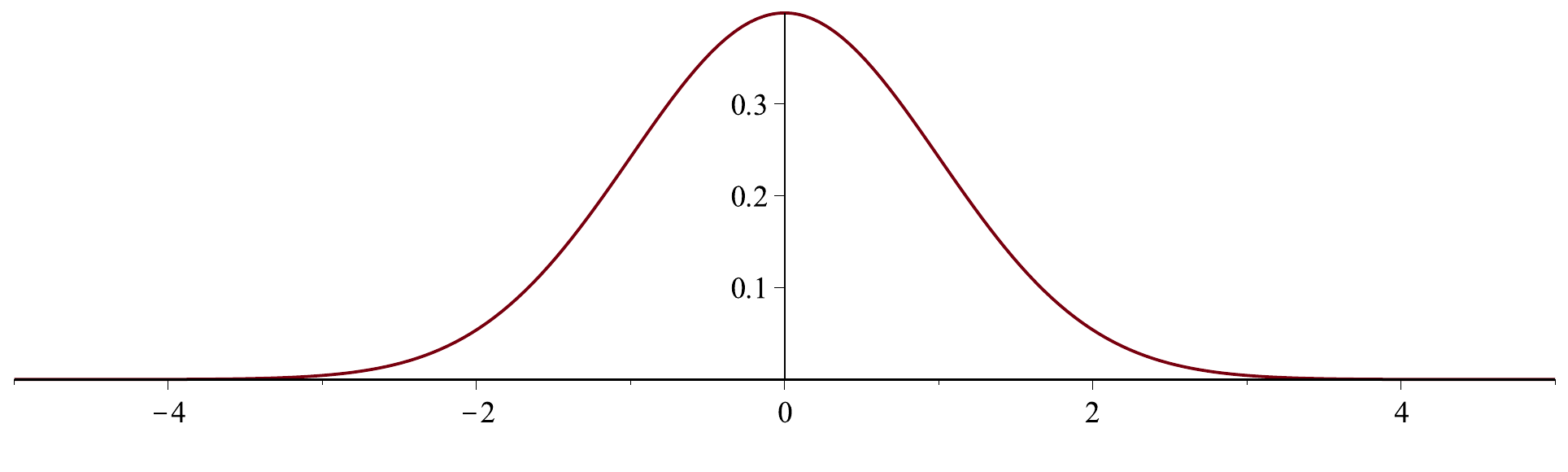}\\
\includegraphics[width=11cm]{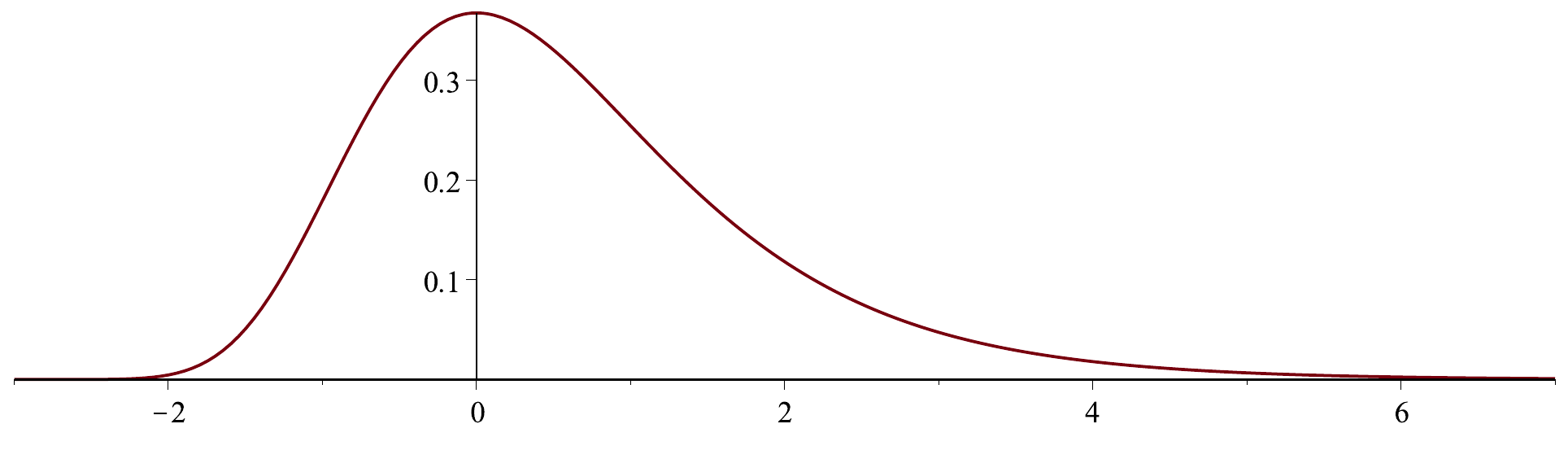}
\end{center}
\caption{\small The densities of Gaussian and Gumbel distributions.}
\label{fig:pdfs}
\end{figure}

 Weak convergence of distributions (also known as convergence in distribution) is denoted by~``$\Rightarrow$''. To state the main mathematical result, we need to recall the standard Gaussian distribution $\Nc$ which has density
 \[
 f_{\Nc}(t)=\frac{e^{-t^2/2}}{\sqrt{2\pi}},\quad t\in\R,
 \]
and the Gumbel distribution~$\Gc$ which has distribution function
\[
F_{\Gc}(t)=e^{-e^{-t}},\quad t\in\R,
\]
and density
\[
f_{\Gc}(t)=e^{-t-e^{-t}},\quad t\in\R.
\]
The densities $f_{\Nc}$ and $f_{\Gc}$ are plotted on Figure~\ref{fig:pdfs}.

\begin{theorem} 
\label{thm:small-noise}
In cases~{\rm\ref{it:right-drift}}, {\rm\ref{it:left-drift}}, and~{\rm\ref{it:barrier}$_1$}, there are constants $A,B>0$
such that
\begin{equation}
\label{eq:CLT}
\Law\left[\frac{\tau_\eps - A}{B\eps}\, \Big|\, \Gamma_\eps \right] \Rightarrow \Nc.
\end{equation}
In case~{\rm\ref{it:barrier}$_0$}, there are constants $A\in\R, B>0$ such that
\begin{equation}
\label{eq:Gumbel-limit}
\Law\left[\,\frac{\displaystyle \tau_\eps - \frac{2}{\lambda}\ln \frac{1}{\eps}- A}{B} \, \Bigg|\, \Gamma_\eps \right] \Rightarrow \Gc.
\end{equation}

\end{theorem}

In other words, conditionally on exit through $R$ (symptom development), in
cases~{\rm\ref{it:right-drift}}, {\rm\ref{it:left-drift}}, and~{\rm\ref{it:barrier}$_1$},
the asymptotic shape of the exit distribution is Gaussian:
\begin{equation}
\label{eq:Gaussian-representation}
\tau_\eps` \stackrel{d}{\approx} A + \eps B N, 
\end{equation}
where $N$ has standard Gaussian distribution, and in case~{\rm\ref{it:barrier}$_0$}, the asymptotic shape of the exit distribution is
Gumbel:
\begin{equation}
\label{eq:Gumbel-limit-representation}
\tau_\eps \stackrel{d}{\approx} \frac{2}{\lambda}\ln \frac{1}{\eps}+A+B G = A_\eps +BG, 
\end{equation}
where $G$ is a Gumbel random variable, and $A_\eps =\frac{2}{\lambda}\ln \frac{1}{\eps}+A$.

We note that although the exit time distribution is right-skewed for all $\eps>0$, the skew asymptotically
vanishes as $\eps\to0$ in cases~{\rm\ref{it:right-drift}}, {\rm\ref{it:left-drift}}, and~{\rm\ref{it:barrier}$_1$}, and there is no contradiction with the symmetry of the limiting Gaussian distribution.

Theorem~\ref{thm:small-noise} in cases~{\rm\ref{it:right-drift}} and~{\rm\ref{it:barrier}$_1$} is a specisfic case of a 
classical result that can be found in~\cite[Section 2.2]{FW2012}. 
For case~{\rm\ref{it:left-drift}}, 
Theorem~\ref{thm:small-noise} was established in~\cite{AB2011a}. 
All these situations can be described as the Levinson case according to the terminology of~\cite{FW2012}.

In case~{\rm\ref{it:barrier}$_0$}, the diffusion trajectories that cross the repelling potential wall at $p$ are often call reactive paths. Theorem~\ref{thm:small-noise} in this case describes the  conditional limit for the length of reactive paths. It was established first in~\cite{ALEA}. For a discussion of these results and other approaches to them, see also \cite{OnGumbel},\cite{Gumbel-preprint},\cite{Berglund:MR3585827}.

We do not give new proofs for any of the cases in Theorem~\ref{thm:small-noise} here. Our contribution is simply reinterpreting these existing results in terms of incubation periods. We discuss this interpretation and broader context in Section~\ref{sec:discussion}. 

\section{Right skewness, positive cumulants: proofs}
\label{sec:proofs-of-positive-skewness}

\subsection{Proof of Theorem~\ref{thm:positive-cumulants} }
In this section we prove Theorem~\ref{thm:positive-cumulants}. The first step is writing down an SDE for the conditioned process.
Conditioned on $\Gamma$, the distribution of process~$X$ coincides with that of the solution of a new SDE 
\begin{equation}
\label{eq:h-transform}
dX(t)=\tilde b(X(t))dt+\sigma(X(t))dW(t).
\end{equation}
Here $\sigma$ is the same is in the original SDE~\eqref{eq:basic_sde}, and
\[
\tilde b(x)=b(x)+\sigma^2(x)\frac{h'(x)}{h(x)},\quad 0<x<R,
\]
where $h(x), x\in[0,R]$ denotes the probability of $\Gamma$ for diffusion~\eqref{eq:basic_sde} started at $x$. 
This is so called Doob's $h$-transform, see~\cite[Section 5]{AB2011a} for the one-dimensional computation and 
\cite[Section 6]{Swiech:MR3461040} for a rigorous and general treatment. 
For all $x\in(0,R)$, we denote by $\Pp_x$ the distribution of the solution of~\eqref{eq:h-transform} with initial condition $X(0)=x$. 
The expectation with respect to $\Pp_x$ is denoted by $\E_x$.
Under our assumptions, all moments of the exit time are finite
for the original equation and thus they are finite for the conditioned one. Moreover, if we define
\[
 \tau_y=\inf\{t\ge 0: X(t)=y\},
\]
then for any $y\in (0,R]$, functions
\[
 \alpha_n(x,y)=\E_x \tau_y^n,\quad n\in \{0\}\cup\N,\quad 0< x\le y,
\]
are smooth in $x\in(0,y]$ up to $y$ and satisfy a hierarchical system of PDEs 
\[
L\alpha_n(x,y)=-n\alpha_{n-1}(x,y), \quad n\in \{0\}\cup\N, \quad 0< x\le y,  
\] 
where $Lf(x)=b(x)f'(x)+\frac{1}{2}\sigma^2(x)f''(x)$ is the generator of the semigroup associated with the diffusion~\eqref{eq:h-transform},
see, e.g., equation (3.38) in~\cite[Chapter 15]{Karlin-Taylor:MR611513}.
Let us also denote the $n$-th cumulant of $\tau_y$ under $\Pp_x$ by $\kappa_n(x,y)$, $0< x\le y\le R$.
Since $\kappa_n(R,R)=0$, we can write
\begin{equation}
\label{eq:integral-rep-for-cumulant}
\kappa_n(x,R)= -(\kappa_n(R,R)-\kappa_n(x,R)) = -\int_x^{R}\frac{d}{dy}\kappa_n(y,R) dy.   
\end{equation}
The strong Markov property implies that under $\Pp_x$, the times $(\tau_y)_{x\le y\le R}$ form a process with independent increments,
and if $0<y_1\le y_2 \le R$,  then the distribution of $\tau_{y_2}-\tau_{y_1}$ under $\Pp_x$ does not depend on $x\in(0,y_1)$. Combining this with
the smoothness of $\kappa_n$, we obtain  
\begin{equation}
\label{eq:cumulant-derivative-at-endpoint}
 \frac{d}{dy}\kappa_n(y,R)=\frac{d^-}{dy}\kappa_n(y,R)= \frac{d^-}{dz}\kappa_n(z,y)\bigg|_{z=y}.
\end{equation}
Using \eqref{eq:cumulant-via-moments} we obtain
\[
\kappa_n(z,y)=\alpha_n(z,y)+P_n(\alpha_1(z,y),\ldots,\alpha_{n-1}(z,y)), 
\]
where  each monomial term constituting $P_n$ is at least of order $2$.  Since $\alpha_1(y,y)=\ldots =\alpha_{n-1}(y,y)=0$, we obtain that the derivative of each of those terms with respect to $z$ at $z=y$
equals~$0$ and thus 
\begin{equation}
\label{eq:nonstrict-inequality}
\frac{d^-}{dz}\kappa_n(z,y)\bigg|_{z=y}= \frac{d^-}{dz}\alpha_n(z,y)\bigg|_{z=y}\le0,  
\end{equation}
where the inequality follows since $\alpha_n(z,y)$ is clearly nonincreasing in~$z$. Let us prove that, in fact, strict inequality
holds:
\begin{equation}
\label{eq:estimate-on-derivative} 
\frac{d^-}{dz}\alpha_n(z,y)\bigg|_{z=y}<0.
\end{equation}
Then the theorem will follow from \eqref{eq:integral-rep-for-cumulant},~\eqref{eq:cumulant-derivative-at-endpoint}, and~\eqref{eq:nonstrict-inequality}.
Let us take any $z_0\in(0,y)$ and 
notice that for $z\in (z_0,y)$
\begin{equation}
\label{eq:lower-estimate-on-f_r}
 \alpha_n(z,y)\ge u(z) \alpha_n(z_0),
\end{equation}
where $u(z)$ denotes the probability that diffusion started at $z$ reaches $z_0$ before $y$. This function satisfies
the equation
\begin{equation}
\label{eq:Fokker-Planck}
b(z)u'(z)+\frac{1}{2}\sigma(z)u''(z)=0,\quad x\in [z_0,y],
\end{equation}
with boundary conditions
\begin{align}
\label{eq:left-boundary-condition} 
u(z_0)&=1,
\\u(y)&=0. 
\label{eq:right-boundary-condition} 
\end{align}
The desired estimate~\eqref{eq:estimate-on-derivative} will follow from \eqref{eq:lower-estimate-on-f_r} if we show that
$u'(y)<0$. Since $u$ is nonnegative and $u(y)=0$, we must have $u'(y)\le 0$. Assuming $u'(y)=0$ would imply, by the uniqueness theorem
for solutions of the regular second-order equation~\eqref{eq:Fokker-Planck}  
and \eqref{eq:right-boundary-condition}, that $u\equiv 0$. The contradiction with~\eqref{eq:left-boundary-condition} shows that $u'(y)<0$
and finishes the proof of the theorem.
\epf

\begin{remark}\rm The theorem and the proof presented here hold in more general situations with minor modifications.
We may have worked with diffusions on $(-\infty,R]$ provided that $\alpha_n(x,R)<\infty$. Assuming the latter condition, the nonstrict 
 inequality~\eqref{eq:nonstrict-inequality} always holds as the proof above shows. For the theorem to hold 
 it is sufficient to have strict inequality at one point $y\in[x,R]$, so we could have required only that
 $\sigma(y)>0$ for some $y\in[x,R]$.
\end{remark}
\begin{remark} \rm
Our soft proof is based on the analysis of the sign of the integrand in~\eqref{eq:integral-rep-for-cumulant}
although  quantitative estimates are also possible. 
\end{remark}

\subsection{Positive cumulants for hitting times in discrete random walks}

In this section, we give a more elementary proof of a version of Theorem~\ref{thm:positive-cumulants} for discrete random walks.

We assume that the evolution $(X_j)_{j\ge 0}$ happens in discrete time on the discrete state 
space~$\N=\{0,1,2,\ldots\}$, it is Markov, time-homogeneous, and nearest neighbor (aka birth-death), i.e., for each $k\in\N$, there is a number $p_k\in(0,1]$ such that if the process is at the site $k$ at time $n$, then at time $n+1$ it jumps to $k+1$ with probability $p_k$ and it jumps to $k-1$ with probability $1-p_k$. We must require $p_1=1$. 
We will denote by $\Pp_k$ the distribution of this process started at $X_0=k$, and $\E_k$ denotes
the expectation with respect to $\Pp_k$. 
If $k,R\in\N$ and satisfy $k< R$,  we denote $\tau_R=\inf\{j\in\N:\ X_j=R\}$.

We note that due to the discrete  Doob's  $h$-transform, this setup automatically contains random walks on $\{0\}\cup\N$ conditioned on reaching~$R$ before~$0$.

\begin{theorem}
\label{thm:right-skew-rw}
Let $k<R$. Then for all $n\in\N$, 
$\kappa_n(\tau)\ge 0$ under $\Pp_k$. The identity in this inequality occurs if and only if $n\ge 2$ and $p_k=p_{k+1}=\ldots=p_{R-1}=1$. 
\end{theorem}

\bpf 
We have
\begin{equation}
\label{eq:representation-via-sum-of-independent}
\tau_R=(\tau_{k+1}-\tau_k)+(\tau_{k+2}-\tau_{k+1})+\dots+(\tau_R-\tau_{R-1}),
\end{equation}
where $\tau_k\stackrel{\as}{=}0$ under $\Pp_k$.
By the strong Markov property, random variables $(\tau_{l+1}-\tau_l)_{l=k}^{R-1}$ are 
independent and the distribution of
$\tau_{l+1}-\tau_l$ (the time it takes to reach $l+1$
starting from $l$)  equals that of $\tau_{l+1}$ under $\Pp_l$.
Since cumulants are additive for sums of independent random variables, it suffices to prove that
cumulants of $\tau_{l+1}$ under $\Pp_l$, i.e.,
the Taylor coefficients of $\ln M_l(\lambda)$ at $0$, are all positive.

Under the conditions of the theorem, for all $l\le R$, the  moment generating function 
\[
M_l(\lambda)= \E_{l} e^{\lambda \tau_{l+1}}  
\]
is well-defined for $\lambda$ in a small neighborhood of $0$.

Under $\Pp_l$, before reaching $l+1$, the process $X$ makes a random number $T\ge 0$ of excursions that involve stepping to $l-1$ first and then after a random number of steps returning to $l$. In other words,
\begin{equation}
\label{eq:representation-of-next-step}
\tau_{l+1}=\sum_{i=1}^T \xi_i+1,
\end{equation}
where 
where $(\xi_{i})_{i\in\N}$ is an i.i.d.\ family independent of~$T$, with a common distribution, that of $\tau_l+1$ 
under $\Pp_{l-1}$. The additional increments of $1$ account for steps from from $l$ to $l-1$ and from $l$ to $l+1$. The distribution of $T$ is geometric:
\begin{equation*}
P_r=\Pp_l\{T=r\}= (1-p_l)^rp_l,\quad r\ge 0.
\end{equation*}
Due to \eqref{eq:representation-of-next-step}, we obtain
\begin{multline*}
M_l(\lambda)=\E_l e^{\lambda \left(\sum_{i=1}^T\xi_i+1\right)} =\sum_{r=0}^\infty p_l(1-p_l)^r 
\E_l   e^{\lambda \left(\sum_{i=1}^j\xi_i+1\right)}
\\ =
p_l e^{\lambda}\sum_{r=0}^\infty (1-p_l)^r 
\left(\E_l   e^{\lambda \xi_i}\right)^r = p_l e^{\lambda}\sum_{r=0}^\infty (1-p_l)^r 
(e^\lambda M_{l-1}(\lambda))^r 
\\ 
= \frac{p_l e^{\lambda}}{1-(1-p_l)e^\lambda M_{l-1}(\lambda)}. 
\end{multline*}

If $p_l=1$, then $M_l(\lambda)\equiv 1$, and $\ln M_l(\lambda)\equiv 0$, so let us consider the situation where $p_l\in(0,1)$.
Since all Taylor coefficients of $e^\lambda$ and $M_{l-1}(\lambda)$ at~$0$ are positive (the latter are the moments of a positive 
random variable), it suffices to check that if a function $f(\cdot)$ satisfies $f(0)=1$ and has all positive Taylor coefficients at $0$, then
for any $q\in(0,1)$ all Taylor coefficients at $0$ of
\[
 g(\lambda)=-\ln (1-qf(\lambda))
\]
are positive. Since the latter directly follows from 
\[
-\ln(1-x)=\sum_{n=1}^\infty \frac{x^n}{n},\quad |x|<1,
\]
the proof is completed.
\epf
\begin{remark}\rm Similarly to the continuous case, one can study random walks that are not bounded below. Then instead
of the finiteness of the moment generating function we might 
only require $\E_l \tau_{l+1}^r<\infty$ and work with charateristic functions
$\phi_l(\lambda)=\E_l e^{i\lambda \tau_{l+1}}=M_l(i\lambda)$
that are defined for all $\lambda\in\R$ and allow for finite order Taylor expansions. Also, it is possible to obtain more
quantitative estimates, a direction that we do not pursue here.
\end{remark}

\subsection{An elementary proof of right-skewness in the discrete random walk case.}

Although the following result is a direct consequence of Theorem~\ref{thm:right-skew-rw} and the definition of skewness, we give
a direct proof that does not use moment generating functions.
\begin{theorem} Suppose $p_l>0$ for all $l$. Then for any $k$ and $R$ satisfying $1\le k \le R$, the distribution of $\tau$ under $\Pp_k$ is right-skewed.
\end{theorem}

\bpf We need to prove that $\kappa_3(\tau_R)\ge 0$. Due to representation~\eqref{eq:representation-via-sum-of-independent}
in terms of a sum of independent hitting times, and since 
cumulants are additive for sums of independent random variables, it suffices to prove that
\begin{equation}
\kappa_{3,\Pp_l}(\tau_{l+1})=\E_l \tau_{l+1}^3-3\E_l \tau_{l+1}^2 \E_l \tau_{l+1}+ 2 (\E_l \tau_{l+1})^3 > 0,\quad l\in \N.
\label{eq:one-step-cumulant-positive}
\end{equation}
We recall the representation~\eqref{eq:representation-of-next-step}.
Since shifts by $1$ do not change the cumulants, we only need to prove the following claim:
if $(\xi_i)_{i\in\N}$ is an i.i.d.\ positive sequence
with $\kappa_3(\xi_1)\ge 0$ and $T$ is an independent geometric variable, then
\begin{equation}
\kappa_3\left(S\right)> 0,
\end{equation}
where $S=\sum_{i=1}^T\xi_i$.

Let $m_r=\E \xi_1^r,$ $r=1,2,3$, and  $p=p_l$ for brevity. Then

\begin{align*}
a_1&= \E T=\frac{1-p}{p},\\
a_2&= \E T(T-1)=\frac{2(1-p)^2}{p^2},\\
a_3&= \E T(T-1)(T-2)=\frac{6(1-p)^3}{p^3},
\end{align*} 
\begin{align*}
\E S &= \sum_{r=1}^{\infty} P_r \E \sum_{i=1}^r \xi_i =  \sum_{r=1}^{\infty} r P_r m_1 = a_1 m_1
=\frac{1-p}{p}m_1,
\\
\E S^2&=  \sum_{r=1}^{\infty} P_r \E \left(\sum_{i=1}^r \xi_i\right)^2= \sum_{r=1}^{\infty} P_r(rm_2+r(r-1)m_1^2)=a_1m_2+a_2m_1^2
\\ & \hspace{7cm} =\frac{1-p}{p} m_2+\frac{2(1-p)^2}{p^2}m_1^2,
\\
\E S^3&=  \sum_{r=1}^{\infty} P_r \E \left(\sum_{i=1}^r \xi_i\right)^3
= \sum_{r=1}^{\infty} P_r(rm_3+3r(r-1)m_1m_2+r(r-1)(r-2)m_1^3)
\\&=a_1m_3+3a_2m_1m_2+a_3m_1^3 = \frac{1-p}{p} m_3+\frac{6(1-p)^2}{p^2} m_1m_2+\frac{6(1-p)^3}{p^3}m_1^3, 
\end{align*}
So
\begin{align*}
\kappa_3(S)=&\E S^3 -3 \E S^2 \E S +2(\E S)^3
\\
=& \frac{1-p}{p} m_3+\frac{6(1-p)^2}{p^2} m_1m_2+\frac{6(1-p)^3}{p^3}m_1^3
\\&-3 \frac{1-p}{p}m_1\left(\frac{1-p}{p} m_2+\frac{2(1-p)^2}{p^2}m_1^2\right)+2 \left(\frac{1-p}{p}\right)^3m_1^3
\\
=&\frac{1-p}{p}  m_3+3\left(\frac{1-p}{p}\right)^2m_1m_2+2 \left(\frac{1-p}{p}\right)^3 m_1^3> 0,
\end{align*}
which completes the proof. \epf

\section{Discussion} \label{sec:discussion}
In this section, we would like to discuss broader context of applicability of our approach as well as its limitations.  

We assumed that the onset of symptoms corresponds to crossing a threshold by a one-dimensional continuous Markov stochastic process. This, of course means, that we are trying to represent the complex process of the propagation of harmful invaders within an organism in the presense of inhomogeneity of tissues, blood circulation, immune response, etc.\ with a single state variable.  
Such a mean field model must be an oversimplification of the reality and cannot possibly be precise.

Also, for a probabilistic model to be useful in applications, one needs certain homogeneity of the data, ideally an i.i.d.\ ensemble to ensure that standard statistical tools based on empirical frequencies and averaging in the law of large numbers are adequate. Assuming that our one-dimensional model gives a fair representation of the dynamics within one infected individual, a better model would account for 
fluctuations in all the parameters due to variability across the population:  the starting point $x_0$, the coefficients $b,\sigma$, 
and the symptom onset level~$R$, especially since in reality the moment of onset of symptoms is defined loosely due to  the symptom
detection dependence on uncontrolled external factors.

It is true that models with more complex joint geometry of the domain and diffusion coefficients, taking
into account non-Markovian effects and variability of parameters
can in principle lead to different behavior of exit times.
However, our conclusions should survive moderate modifications of the model and be applicable
for a broad class of stochastic models. For example, if the parameters of the model can vary and    
 form a statistically homogeneous ensemble, the exit distribution under small noise will be then described by~\eqref{eq:Gaussian-representation} or~\eqref{eq:Gumbel-limit-representation} with random 
values of $A$ and $B$, i.e., this is a weighted mixture of a family of Gaussian- or Gumbel-shaped distributions.
Assuming that the fluctuations of the parameters are small, the shape of the distribution will still be very similar to Gaussian or Gumbel.

Our results for limiting shapes of exit time distributions are obtained in the limit of small noise. Although smallness of the noise is a natural assumption, it is not a~priori obvious that it holds in reality. One can say though that the agreement of the real data with
Gumbel distribution reported in~\cite{Strogatz}
(see Figure~\ref{fig:data-1949-and-1950}) is an argument in the favor of small noise hypothesis in
case~III$_0$, with a repelling critical point between the staring point and the symptom onset level~$R$.  

If the noise is not small, then our results show that the exit distribution for our model has right skew but precise computations of exit time or conditional exit time distributions
become hard.
 In general, the computations can be based on solving second order differential equations for characteristic functions or moment  generating functions, see~\cite{ALEA} or numerical simulations. Since all these distributions have right skew, it is difficult to distinguish between them, so one and the same data set can be equally well approximated by Gumbel or lognormal density. This point seems to be mentioned for the first time in literature in~\cite{Strogatz}.  

\begin{figure}
\begin{center}
\includegraphics[width=10cm]{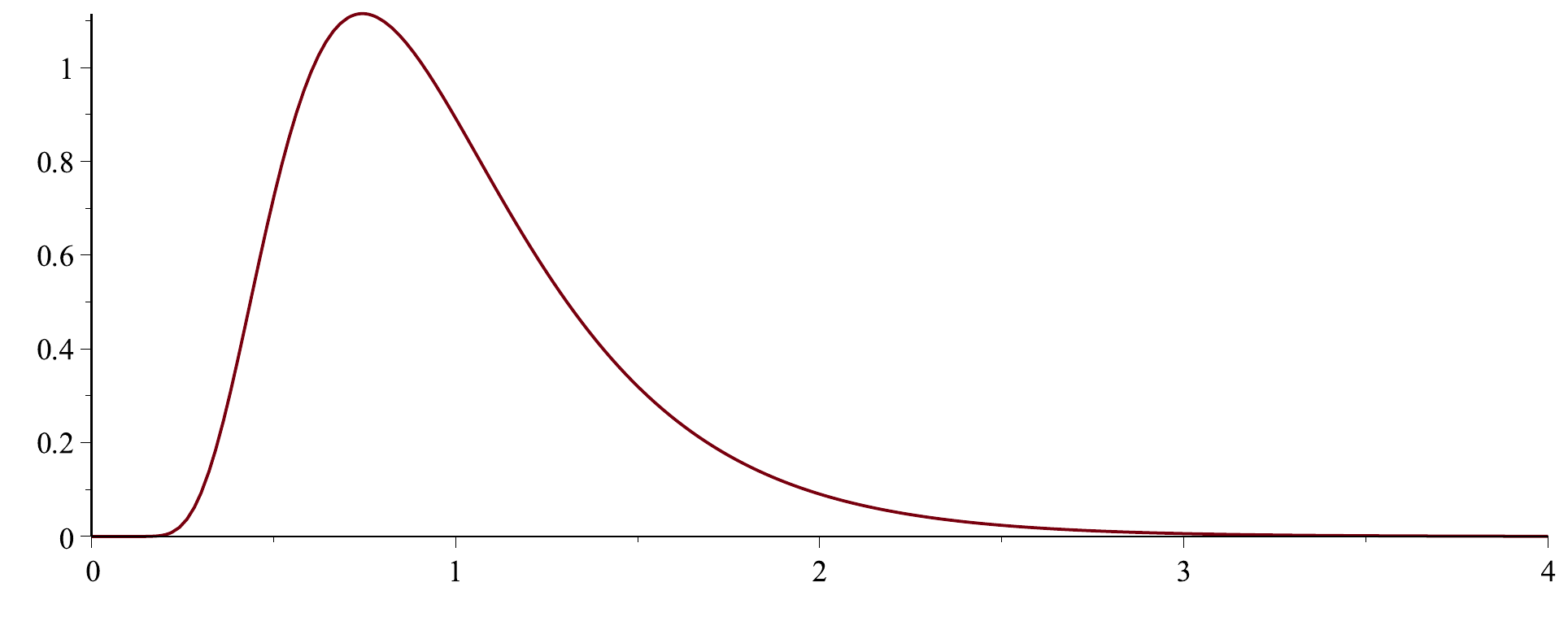}\\
\includegraphics[width=10cm]{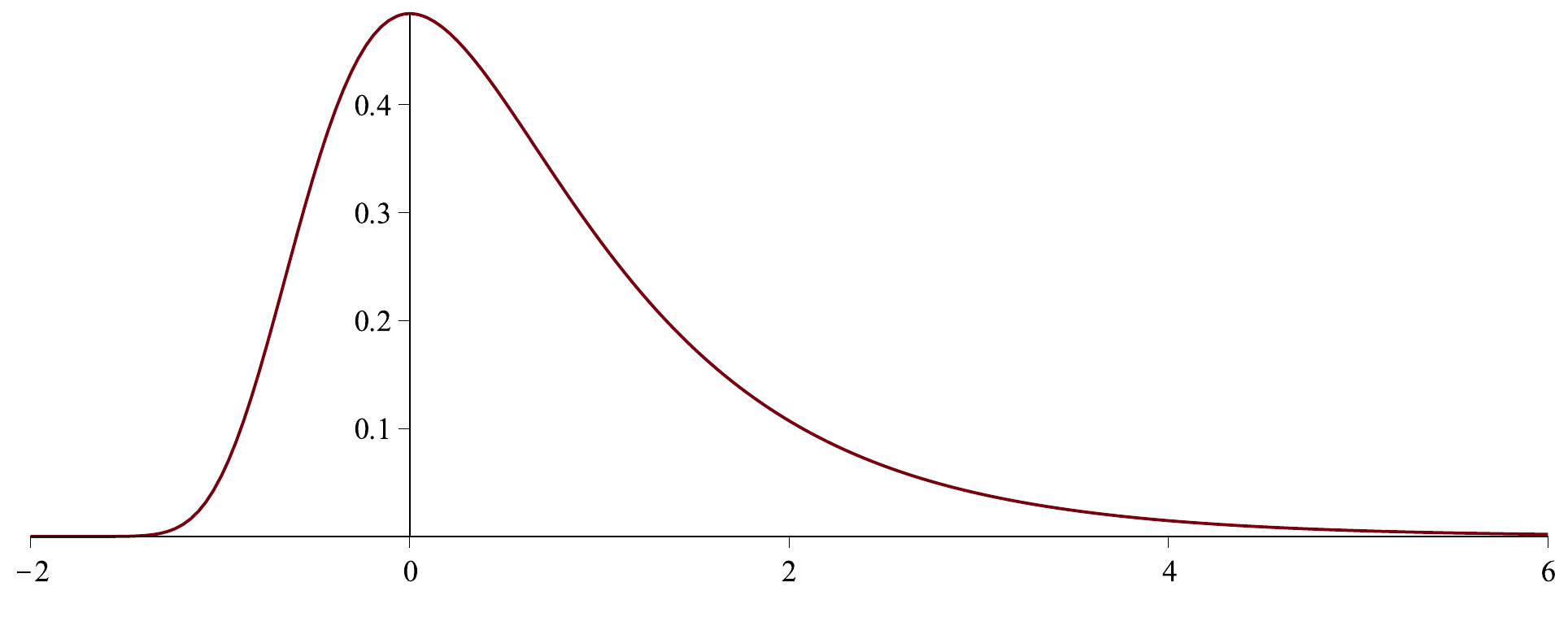}
\end{center}
\caption{\small $(1,5)$-Inverse Gaussian and exp-Gaussian densities}
\label{fig:inverse-g}
\end{figure}

There are few other situations where exit time distributions or their limits are known. One is the one-sided exit problem    
on  $(-\infty,R]$ for Brownian motion with nonnegative drift, where the drift and diffusion coefficients are both constant, $b\ge 0$ and $\sigma>0$. The exit time distribution 
(first obtained in~\cite{Schrodinger}) is known to be  Wald, or Inverse Gaussian $\Ic(R/b, R^2/\sigma^2)$, where $\Ic(\mu,\lambda)$
stands for the distribution with density 
\[
f_{\Ic(\mu,\lambda)}(t)=\left(\frac{\lambda}{2 \pi t^3}\right)^{1/2} \exp\left\{\frac{-\lambda (t-\mu)^2}{2 \mu^2 t}\right\},
\]
extended by continuity to  $\mu=+\infty$ to include the case $b=0$, see, e.g., (73)--(74) in~\cite{Cox-Miller:MR0192521}. Figure~\ref{fig:inverse-g} plots the density of~$\Ic(1,5)$.

The limiting shape of the distribution for conditional exit time in case~III with initial condition 
at the critical point $x_0=p$ was first computed in~\cite{Day:MR1110156}.
It is the distribution of  $-\ln|N|$,
where $N$ is a standard Gaussian random variable, and thus
can be called exp-Gaussian, $\Ec$:
\begin{equation*}
f_{\Ec}(t)=\sqrt{2}{\pi}e^{\textstyle -t-\frac{ e^{-2t}}{2}},
\end{equation*}
see~Figure~\ref{fig:inverse-g}. The universality  of this distribution and its generalizations 
was studied in~\cite{Bakhtin-SPA:MR2411523},\cite{Bak2010},\cite{Bak2011},\cite{Gumbel-preprint},%
\cite{Bakhtin-Correll:MR2983392}, \cite{Bakhtin:MR2935120}.

In fact, in the latter two papers, the emergence of this distribution in decision-making in humans
and in associated diffusion models and discrete agent-based neuronal models with mean-field
interactions of Curie--Weiss type was studied. The decision/reaction/response times have been studied in
psychological literature for more than a century, 
 and one of the natural approaches is to use diffusions to model these times.
see the bibliography in~\cite{Luce-book}
and~\cite{Bakhtin-Correll:MR2983392}. Although it has been observed that most response times data are right-skewed, that fact has not received mathematical justification until the present work. 
We believe that the present paper proving that
exit times of Markov diffusions conditioned on the direction of exit are always right-skewed provides a simple and
robust answer to this question. Our result can be used as a test for applicability of
diffusion models of this kind: if the data are left-skewed, then no $1$-dimensional diffusion model can reproduce 
these data.

It is worth mentioning 
that~\cite{Bakhtin-Correll:MR2983392}  and the present paper
contain the first mathematically rigorous results on distributions of 
response times understood as exit times. What sets this work apart from the existing literature
besides the mathematical
rigor is that we are able to make a universality claim: we show that certain 
features of random variables involved must hold for a broad class of models.

In~\cite{Bakhtin-Correll:MR2983392}, it 
is the universality of the shapes of decision making times in symmetric decision tasks with
no a~priori bias in small noise situations. In the present paper, it is the universality of right-skewness of the distributions of exit times and their limiting Gaussian or Gumbel 
(depending on the macroscopic robust features of the phase portrait) shapes in asymmetric small noise situations.

One reason of the universal behavior in our works comes from modeling with SDEs.  Their big advantage
(in comparison with models of the kind considered in~\cite{Strogatz},\cite{Strogatz:PhysRevE.96.012313}) is that each SDE defines a whole universality class such that the macroscopic behavior of the models
in the class can be effectively described by the exemplar SDE.  This includes discrete and continuous random dynamics. The classical examples of this are the Gaussian limit in the Central Limit Theorem and Donsker's Invariance Principle 
stating that random walks with i.i.d.\ increments and finite variance are in the universality class of the Wiener process, the simplest disffusion, see, e.g.,
\cite[Chapter~7]{Ethier-Kurtz:MR838085}. Useful examples of such limit theorems abound in the literature, and here we mention just 
one: the reason why the exp-Gaussian distribution shape appears as the limiting one for the discrete
Curie--Weiss model of neuronal interaction in \cite{Bakhtin-Correll:MR2983392} 
and~\cite{Bakhtin:MR2935120} is that
the model belongs to the universality class of a diffusion near an unstable critical point. It is this universality that allows us to conjecture that statistical features discussed in~\cite{Bakhtin-Correll:MR2983392} and this paper will
be discovered in many other situations. 

A striking example of universality is the Gumbel distribution which appears as the universal
limit in at least three domains: (i)~theory of extreme values, (ii)~theory of residual lifetimes, and (iii)~theory of exit times. Although Gumbel (or double exponential) distribution appears in~\cite{Luce-book} along with a dozen of other distributions that resemble many response time data sets, no convincing explanation of its relevance is given there. The common roots of emergence of
the Gumbel distribution in (i),(ii), and (iii) are discussed in~\cite{Gumbel-preprint}.

In the end of this discussion, let us empasize that the problem of the universal statistical behavior of halting or decision times is very broad. For an example of a seemingly totally different nature, such universal behavior has been observed in halting times for several algoritms and massive random initial data sampled
from various basic ensembles used in mathematical physics in
\cite{Deift:MR3723332},\cite{Deift:MR3380694},\cite{Deift:MR3276499},\cite{Sagun-Trogdon-LeCun}. 
It was rigorously
established in a special case in~\cite{Deift-Trogdon:Toda}.
Although various detection/halting/decision/hitting times appear to belong to different universality classes, this body of observations calls for further study of the universality phenomena for time statistics  in various contexts.

{\bf Acknowledgments.} The author is grateful to Charles Peskin and Percy Deift for 
bringing~\cite{Strogatz} to his attention. He also thanks them for stimulating discussions and
encouragement.

\bibliographystyle{alpha}
\bibliography{happydle}

\end{document}